\documentclass
[aps,preprint,showkeys,a4paper,tightenlines,superscriptaddress,12pt]{revtex4}%
\usepackage{eurosym}
\usepackage{amsfonts}
\usepackage{amsmath}
\usepackage{amssymb}
\usepackage{graphicx}
\usepackage{subfig}
\usepackage{caption}
\usepackage{setspace}%
\providecommand{\U}[1]{\protect\rule{.1in}{.1in}}
\providecommand{\U}[1]{\protect\rule{.1in}{.1in}}
\newtheorem{theorem}{Theorem}
\newtheorem{acknowledgement}[theorem]{Acknowledgement}

\begin{document}
\title{On stickiness of multiscale randomly rough surfaces}
\author{G. Violano}
\affiliation{Department of Mechanics, Mathematics and Management, Politecnico of Bari, V.le
Japigia, 182, 70126, Bari, Italy}
\author{L.Afferrante}
\affiliation{Department of Mechanics, Mathematics and Management, Politecnico of Bari, V.le
Japigia, 182, 70126, Bari, Italy}
\author{A.Papangelo}
\affiliation{Department of Mechanics, Mathematics and Management, Politecnico of Bari, V.le
Japigia, 182, 70126, Bari, Italy}
\affiliation{Hamburg University of Technology, Department of Mechanical Engineering, Am
Schwarzenberg-Campus 1, 21073 Hamburg, Germany}
\author{M. Ciavarella}
\affiliation{Department of Mechanics, Mathematics and Management, Politecnico of Bari, V.le
Japigia, 182, 70126, Bari, Italy}
\affiliation{Hamburg University of Technology, Department of Mechanical Engineering, Am
Schwarzenberg-Campus 1, 21073 Hamburg, Germany}
\email{[Corresponding author. ]Email: michele.ciavarella@poliba.it}
\keywords{Roughness, Adhesion, Dalhquist criterion, DMT model}
\begin{abstract}
We derive a very simple and effective stickiness criterion for solids having
random roughness using a new asymptotic theory, which we validate with that of
Persson and Scaraggi and independent numerical experiments. Previous claims
that stickiness may depend on small scale quantities such as rms slopes and/or
curvatures, obtained by making oversimplified assumptions on the contact area
geometry, are largely incorrect, as the truncation of the PSD spectrum of
roughness at short wavelengths is irrelevant. We find stickiness is destroyed
typically at roughness amplitudes up to three orders of magnitude larger than
the range of attractive forces. With typical nanometer values of the latter,
the criterion gives justification to the qualitative well known empirical
Dalhquist criterion for stickiness which demands adhesives to have elastic
modulus lower than about 1MPa. The results clarifies a much debated question
in both the scientific and technological world of adhesion, and may serve as
benchmark for better comprehension of the role of roughness.

\end{abstract}
\maketitle

\section{Introduction}

Contact mechanics with roughness has made tremendous progress in recent years,
and adhesion has become increasingly relevant with the interest on soft
materials, nano-systems and the analysis of bio-attachments (for two recent
reviews \cite{Ciavarella2018adh,Vakis2018}). Contact between solids occurs via
large van der Waals forces, usually represented, for example, by the well
known Lennard-Jones force-separation law. These force give rise to a
theoretical strength much higher than the typical values to break bulk
materials apart.\ Hence, it appears as an "adhesion paradox"
\cite{Kendall2001} that all objects in the Universe should stick to each
other. This does not happen due to inevitable surface roughness at the
interface, and Nature has developed different strategies to achieve
neverthless robust stickiness, including contact splitting and hierarchical
structures (\cite{Autumn, Gao1, Gao2}). At macroscale, it appears that the
only solution to mantain stickiness is to reduce the elastic modulus. This is
well known in the world of Pressure-Sensitive Adhesives (PSA), soft polymeric
materials showing instantaneous adhesion on most surfaces upon application of
just a light pressure \cite{Creton1996}. Dahlquist \cite{Dahlquist1969,
Dahlquist} proposed that to achieve a universal stickiness, the elastic Young
modulus should be smaller than about 1MPa (at 1Hz, as adhesive are strongly
viscoelastic, their modulus depends on frequency). This criterion has no
scientific validation, but appears to be largely used in the world of adhesives.

There have been various attempts to study the problem of elastic contact with
roughness and adhesion. Fuller and Tabor (FT, \cite{FT}) used the Greenwood
and Williamson \cite{GW} concept of describing a rough surface with a
statistical distribution of identical asperities of radius $R$, together with
JKR theory for the sphere contact \cite{JKR}. FT found that adhesion was
easily destroyed with RMS amplitude of roughness $h_{\mathrm{rms}}$ of a few
micrometers in spherical rubber bodies against rough hard plastic surfaces.
Their theory depends only on a single dimensionless parameter $\theta
_{FT}=h_{\mathrm{rms}}^{3/2}\Delta\gamma/\left(  R^{1/2}E^{\ast}\right)  $
where $E^{\ast}$ is the plane strain elastic modulus, $\Delta\gamma$ is
interface energy. The choice of $R$ seems critical in view of its sensitivity
to "resolution" or "magnification" \cite{Ciavarella2017b}, i.e. on the
shortest wavelength in the roughness spectrum. In the "fractal limit", i.e.
for an infinite number of wavelength $R\rightarrow0$, there would be
\textit{no stickiness for any surface}, irrespective of the geometrical
characteristics, like fractal dimension, or root mean square heigths (rms)
amplitude. Hence, FT apparent good correlation with the theory despite the
many limitations (see \cite{Greenwood2017}), may have been due to a fortuitous
choice of $R$ at a relatively coarse scale where measurements were made at
that time.

The JKR theory is not appropriate when contact spots become very small, and
another theory is more promising in this case, which takes the name of DMT for
the case of the sphere \cite{Derjaguin1975}. It makes it possible to solve
contact problems with adhesion using results from the adhesionless problem, by
assuming that the adhesive stresses do not alter the pressure in the contact
area (which therefore remains purely under compression, and remains defined in
the same way as "repulsive") nor the gaps outside the contact. The external
pressure is therefore the difference of the repulsive and an adhesive pressure
$p_{ext}=p_{rep}-p_{ad}$.

Pastewka \& Robbins (PR, \cite{PR2014}) presented a criterion for adhesion
between randomly rough surfaces after interpreting simulations of adhesive
rough contact with relatively narrow band of roughness (as limited by present
computational capabilities, see the Contact Mechanics Challenge of Muser
\textit{et al. }\cite{MuserCC}), i.e. with wavelength from subnanometer to
micrometer scales. Defining "magnification" as the ratio $\zeta=q_{1}/q_{0}$
between the high \ $q_{1}$ and the low $q_{0}$ truncating wavevectors defining
the spectrum of roughness, this means from $\zeta\sim100$ and up to $\zeta
\sim1000$. PR attempted to interpret the results on the (repulsive) area vs
(external) load slope on the basis of a simplified DMT-like model using only
the asymptotic expression for gaps at the edge of cylindrical regions defining
a `boundary layer' surrounding the `repulsive' contact zone, which we shall
here generalize removing some of the strong assumptions in the original
derivation. PR obtained then a criterion that depends mainly on local slopes
and curvature, i.e. on the tail of the PSD spectrum, and leads to a condition
for stickiness $\zeta^{-4+5D/3}<c$, where $c>0$. This implies all surfaces
with fractal dimension $D<2.4$ should be \textit{always} sticky in the fractal
limit. This conclusion seems quite counterintuitive, and in a sense
\textit{more paradoxical} than the Fuller and Tabor one, as most natural
surfaces and surfaces of engineering interest, e.g., polished or sandblasted,
are indeed fractals over a wide range of scales and with $D<2.4$
\cite{Persson2014}. So, this, again, would lead to the "sticky Universe" of
Kendall \cite{Kendall2001}. Other recent theories by Ciavarella
\cite{Ciavarella2018} with the so-called BAM model (Bearing Area Model) and
Joe, Toughless and Barber (JTB, \cite{JTB2018}) seem less paradoxical as
estimate the pressure at pull-off between surfaces does not depend much on
local slopes and curvature.

In order to derive a better criterion than FT an PR, it becomes imperative to
dispose of a theory with enough accuracy for very broad spectra, typical of
real surfaces which can be expected to have features from millimeter to
nanometer scale, hence showing perhaps five decades of roughness wavelengths,
or more. Persson and Scaraggi (PS theory, \cite{PS2014}) provided a full
theory based on the DMT assumption which, with some refinements \cite{Aff2018}%
, can be used for the purpose of deriving a "stickiness" criterion and its
convergence in the fractal limit, finding the correct parameters dependences.

\section{A DMT theory based criterion}

The Lennard-Jones force-separation law is usually represented as%
\begin{equation}
\sigma_{ad}(u)=\frac{8\Delta\gamma}{3\epsilon}\left[  \frac{\epsilon^{3}%
}{g^{3}}-\frac{\epsilon^{9}}{g^{9}}\right]  \label{LJ}%
\end{equation}
where $\Delta\gamma=\int_{\epsilon}^{\infty}\sigma(g)dg$ is the
\textit{interface energy} or, by definition, the work done in separating two
bodies from the equilibrium position $g=\epsilon$, at which $\sigma_{ad}=0$,
per unit area of interface. The maximum tensile traction happens at a
separation $g=3^{1/6}\epsilon$ and is $\sigma_{th}=16\Delta\gamma/\left(
9\sqrt{3}\right)  \epsilon$. A possible simplification is to use a constant
force-law \cite{Maugis1992}.\ Considering gaps from the equilibrium point
namely $u=g-\epsilon$, imposing the same interface energy $\Delta\gamma$,%
\begin{equation}%
\begin{tabular}
[c]{ll}%
$\sigma_{ad}\left(  u\right)  =\sigma_{0},$ & $u\leq\epsilon$\\
$\sigma_{ad}\left(  u\right)  =0,$ & $u>\epsilon$%
\end{tabular}
\ \label{maugis}%
\end{equation}
where $\sigma_{0}=9\sqrt{3}\sigma_{th}/16\simeq\sigma_{th}$ and $\epsilon$ is
the same range of attraction, so obviously $\Delta\gamma=\sigma_{0}\epsilon$.

Notice that if $E^{\ast}$ is the plane strain elastic modulus, $l_{a}%
=\Delta\gamma/E^{\ast}$ defines a characteristic adhesion length which for the
typical Lennard Jones description of an interface between crystals of the same
material is $l_{a}\simeq0.05\epsilon$. The theoretical strength in this case,
$\sigma_{0}=l_{a}E^{\ast}/\epsilon=0.05E^{\ast}$ represents a very high value.

In DMT theories, the adhesive pressure is computed by convolution of the
elementary tension-separation law $\sigma_{ad}\left(  u\right)  $ with the
distribution of gaps $P\left(  u\right)  $, which for the Maugis potential
(\ref{maugis}), simplifies to%
\begin{equation}
p_{ad}=\sigma_{0}\int_{0}^{\epsilon}duP\left(  u\right)  =\sigma_{0}%
\frac{A_{ad}}{A_{nom}} \label{attractive}%
\end{equation}
where $A_{ad}$ is tha "adhesive" contact area, i.e. the region where tensile
stress are applied, and $A_{nom}$ is the "nominal" or "apparent" contact area.
An elaborate expression for $P\left(  u\right)  $ (for the purely repulsive
problem, i.e. in the absence of any adhesion) is obtained in Persson's theory
(see \cite{Aff2018}).

\begin{figure}[ptbh]
\begin{center}
\includegraphics[width=17.0cm]{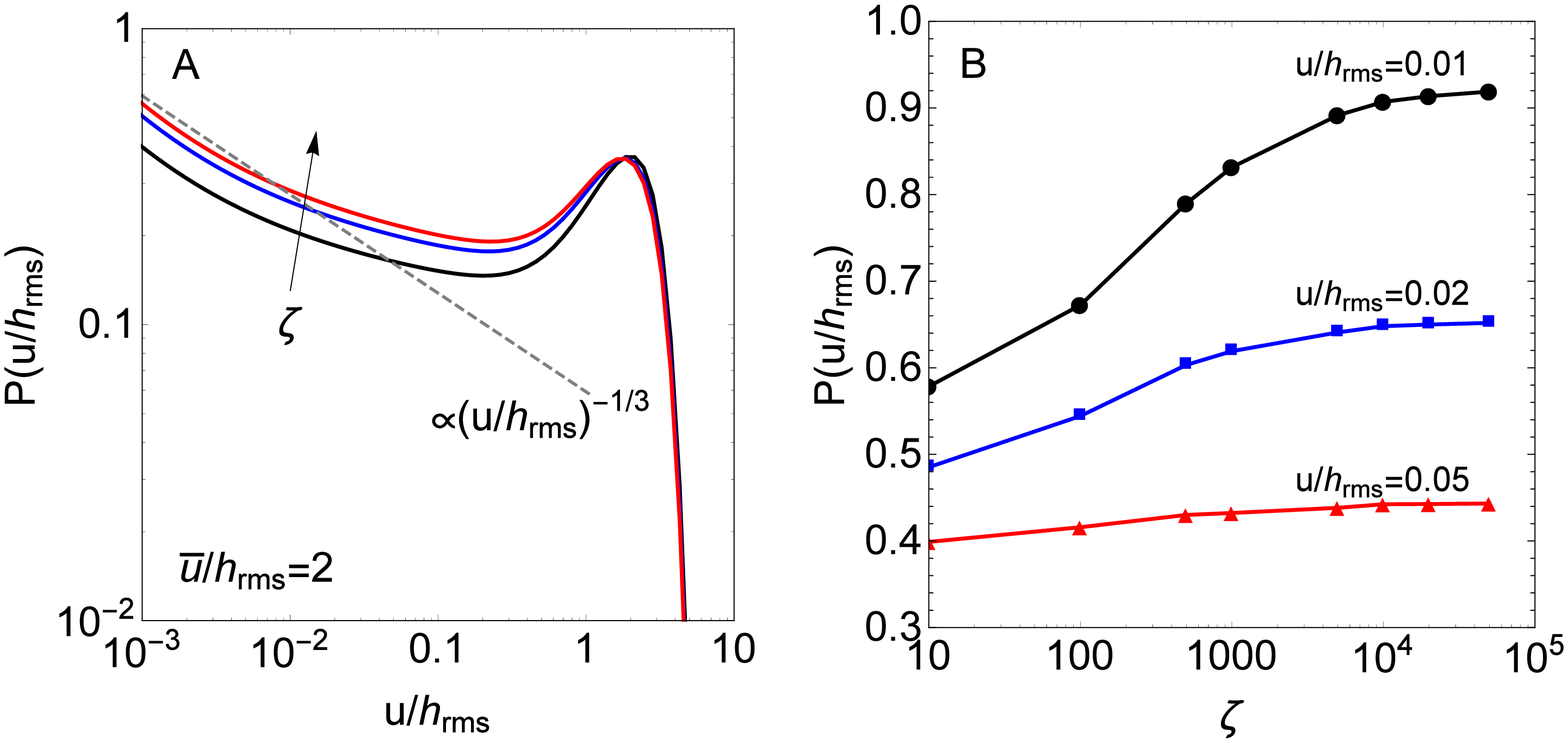}
\end{center}
\caption{(A) Distribution of gaps $P\left(  u\right)  $ (black, blue and red
line respectively for $\zeta=10,$ $100,$ $1000$) with an asymptotic fit
$P\left(  u\right)  \sim u^{-1/3}$ as guide to the eye (dashed black line).
Case with fractal dimension $D=2.2$ (or Hurst exponent $H=0.8$). interfacial
mean separation $\overline{u}/{h}_{rms}=2$ and pure power law power spectrum
random roughness. (B) Convergence of $P\left(  u\right)  $ with increasing
magnifications for various values of gap $u/h_{rms}$. Case with $H=0.8,$
$\overline{u}/{h}_{rms}=1.37$.}%
\label{fig1}%
\end{figure}

Results of the stickiness criterion of PR seem to suggest that there should be
no asymptotic expression (in the fractal limit) to $P\left(  u\right)  $.
However, Fig. \ref{fig1} shows that there is a convergence in the distribution
for increasing magnification $\zeta$ and furthermore there is an asymptotic
scaling at low separations $P\left(  u\right)  \sim u^{-1/3}$. Results in Fig.
(\ref{fig1}a) are given for pure power law PSD (Power Spectrum Density),
fractal dimension $D=2.2$ (i.e. Hurst exponent $H=3-D=0.8$) and for a mean gap
$\overline{u}$ equal to twice the RMS amplitude of roughness, $\overline
{u}/{h}_{rms}=2$. We define a non-dimensional pressure $\widehat{p}%
_{rep}=p_{rep}/\left(  E^{\ast}q_{0}h_{rms}\right)  $ and we remark that, for
typical real surfaces $H\gtrsim0.6$, in the limit of relatively large $\zeta$
and small pressures, Persson's theory reduces to $\widehat{p}_{rep}\simeq
\exp\left(  -2\overline{u}/{h}_{rms}\right)  $ (\cite{Persson2007,Pap2017}).
As we are essentially interested in the region of the area-load relationship
near the axes origin, we disregard $\overline{u}/{h}_{rms}<1$, and also
$\overline{u}/{h}_{rms}>3$ where we are likely to have finite effects due to
poor statistics of the Gaussian surfaces and very few asperities in contact.
This corresponds therefore to the range $\widehat{p}_{rep}=10^{-3}\div10^{-1}$.

Furthermore, the range of attractive forces of interest is $\epsilon<<h_{rms}%
$, where we can assume that the main contribution to the gaps and hence to
adhesion comes from the asymptotic value of $P\left(  u\right)  $ at low $u$,
namely from the regions close to the contact boundaries. We can obtain this
asymptotic form of $P\left(  u\right)  $ from standard contact mechanics
theory, and from the asymptotic part of Persson's theory \cite{Persson2001},
whereas we shall use the full Persson's theory only for the actual calculation
of the prefactors (see Methods).

Specifically, as in detail shown in Methods, the attractive area can be given
as
\begin{equation}
\frac{A_{ad}}{A_{nom}}=\frac{3}{2}a_{V}\widehat{p}_{rep}\left(  \frac
{\epsilon}{h_{rms}}\right)  ^{2/3} \label{area-new-formula}%
\end{equation}
being the coefficient $a_{V}$
\begin{equation}
a_{V}=\frac{3}{2}\left(  \frac{4}{9}\right)  ^{1/3}\frac{9}{32\beta^{8/3}%
}\frac{\left\langle d^{1/3}\right\rangle }{\left\langle d\right\rangle
}E^{\ast3}\sqrt{\frac{2}{\pi V^{3}}}q_{0}h_{rms}^{5/3} \label{a(V)}%
\end{equation}
where $V=\frac{1}{2}E^{\ast2}m_{2}$ is the variance of full contact pressures,
$m_{2}$ is the mean square profile slope along any direction (for a isotropic
surface), $d$ is a\ local characteristic length scale, $\beta$ is a
geometrical factor of order 1, and $\left\langle \cdot\right\rangle $ denotes
the mean value of the argument. Notice, in eqt. (\ref{area-new-formula}),
$A_{ad}/A_{nom}$ is proportional to the external mean pressure if\textit{\ }%
$\left\langle d^{1/3}\right\rangle /\left\langle d\right\rangle $%
\textit{\ }does not depend on pressure.

It is not worth to compute $a_{V}$ directly from (\ref{a(V)}) because of the
problematic term $\left\langle d^{1/3}\right\rangle /\left\langle
d\right\rangle $, and therefore we fit the global $A_{ad}/A_{nom}$ vs.
$\left(  \epsilon/h_{rms}\right)  ^{2/3}$ curves obtained from numerical
results of the adhesionless contact problem with a BEM code, or with the full
Persson's theory.

\begin{figure}[ptbh]
\begin{center}
\includegraphics[width=17.0cm]{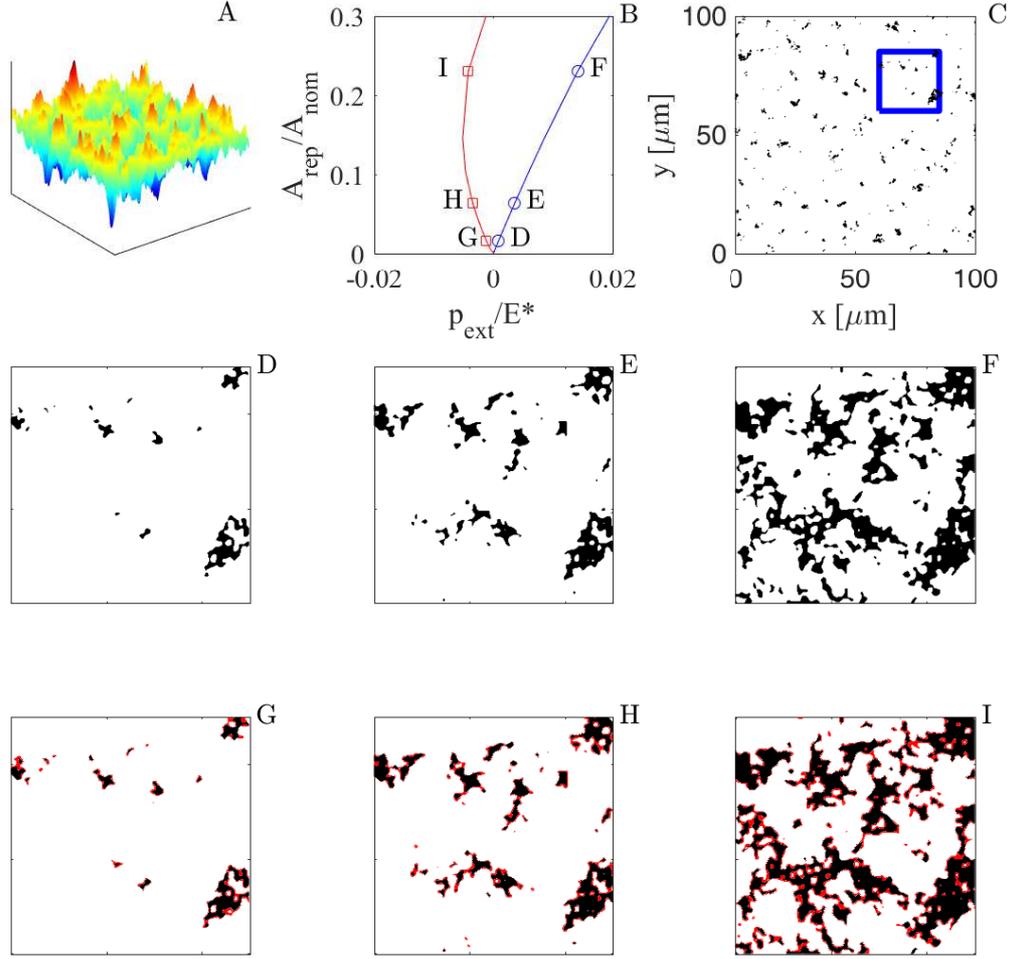}
\end{center}
\caption{Geometry of surfaces and contacting regions. (a) Example of
self-affine fractal surface with $D=2.2$. (b) Dependence of the relative
contact area $A_{rep}/A_{nom}$ on the normalized external pressure
$p_{ext}/E^{\ast}$ as obtained by the numerical BEM code of Pastewka
\cite{Pastewka2012} for the adhesionless and the adhesive case. (D-F)
Evolution of contact spots area for various loads in the linear range and in
absence of adhesion as obtained by BEM. (G-I) Evolution of contact spots area
for various loads in presence of adhesion ($l_{a}/\epsilon=0.2,$
$\epsilon/h_{rms}=7.5\times10^{-3}$) as obtained by BEM.}%
\label{fig2}%
\end{figure}

Fig. \ref{fig2} shows, for a self-affine fractal surface, the repulsive area
vs. external pressure predictions as obtained by the BEM code of Pastewka
\cite{website, Pastewka2012} and the modified Persson's theory given by
Afferrante \textit{et al.} \cite{Aff2018}. It is clear that for nonadhesive
contact the area vs. load relation is approximately linear with a coefficient
$\kappa\simeq2$ (as shown in \cite{Putignano2012,Putignano2012b}). However,
increasing $l_{a}/\epsilon$ the area increases more rapidly with load and a
threshold value of $l_{a}/\epsilon$ exists above which the slope $\kappa$
becomes negative (Fig. \ref{fig2}b) and a nonzero pull-off force exists.
Notice that under load control, upon approach, surfaces are expected to jump
into contact resulting in a finite area of contact, and this may explain why
PR obtained the slope never higher than the vertical.\ In principle,
"displacement control" is not well defined for infinite surfaces which have no
defined stiffness.

Furthermore, in Fig. \ref{fig2} the contacting regions (in black/red the
repulsive/adhesive contact area) as obtained with BEM are given with (Fig.
\ref{fig2}G-H-I) and without adhesion (Fig. \ref{fig2}D-E-F). It is clear that
at increasing pressures contacts become larger and new contacts appear, but
even to the eye, there is no constant representative diameter of contact.

\begin{figure}[ptbh]
\begin{center}
\includegraphics[width=17.0cm]{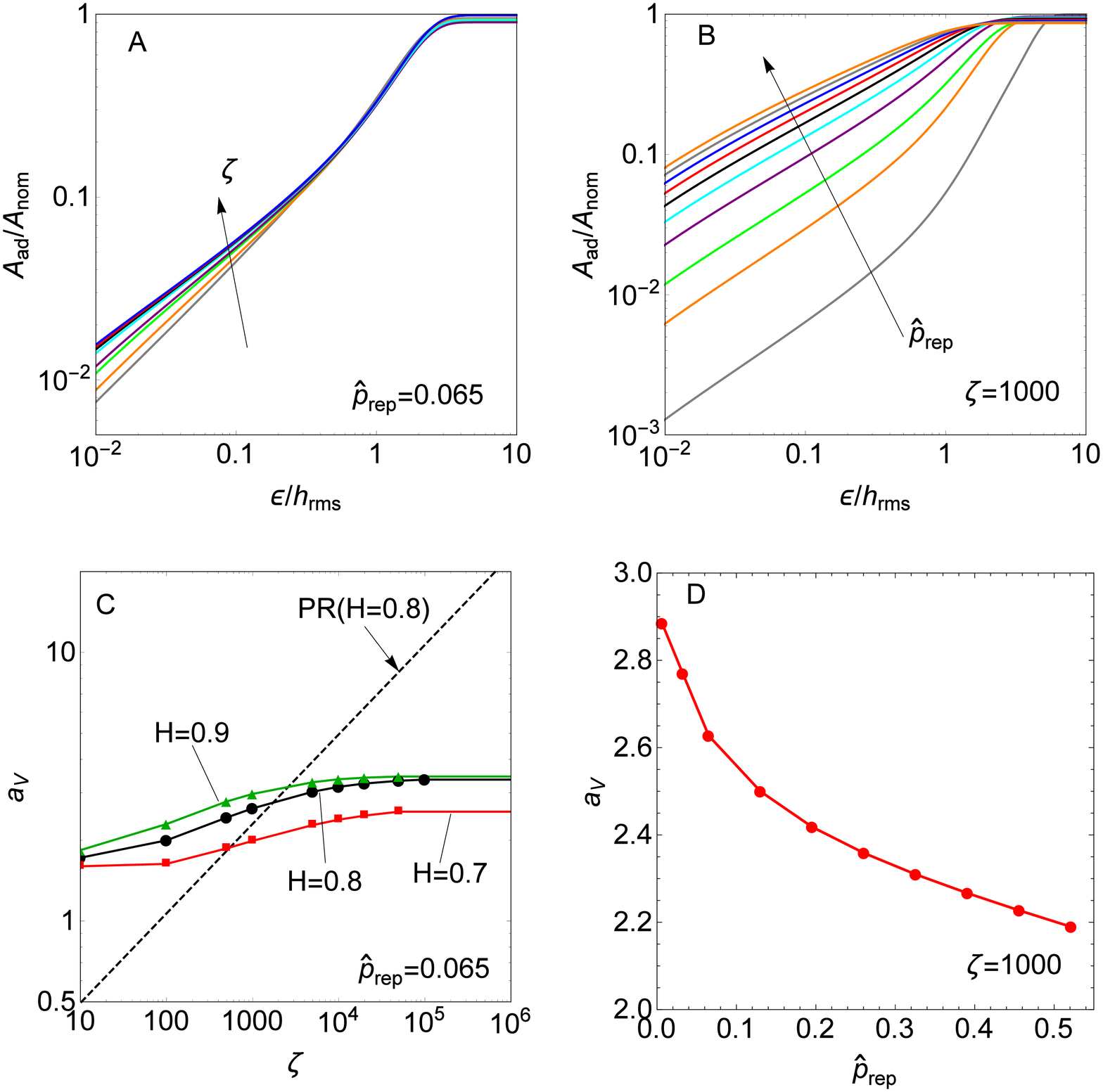}
\end{center}
\caption{(A) Comparison of the attractive area $A_{ad}/A_{nom}$ estimated by
Persson's theory (with $\zeta=\left[
10,100,500,1000,5000,10000,20000,50000\right]  ,$ $H=0.8$) . (B) Comparison of
the attractive area $A_{ad}/A_{nom}$ estimated by Persson's theory (with
$p_{rep}=\left[  6.5\ast10^{-3}%
,0.03,0.07,0.13,0.20,0.26,0.33,0.39,0.46,0.52\right]  $ and $\zeta=1000$). (C)
Estimates of the prefactor $a_{V}$ from Persson's theory for $H=[0.7-0.8-0.9]$
vs PR estimate ($H=0.8$) (black dashed line, $\left[  a_{V}\right]
_{PR}=0.252\zeta^{1/3}$ as obtained in Supp. Information), as a function of
magnification $\zeta$ for $\widehat{p}_{rep}=0.065$. (D) Estimates of the
prefactor $a_{V}$ for $\zeta=1000$ from Persson's theory as a function of for
$\widehat{p}_{rep}$.}%
\label{fig3}%
\end{figure}

Fig. \ref{fig3} shows that in terms of actual adhesive area, the convergence
with $\zeta\ $is very rapid (Fig. \ref{fig3}A) and is not modified by the load
(Fig. \ref{fig3}B). Accordingly, the prefactor $a_{V}$ rapidly converges with
magnification (Fig. \ref{fig3}C), and weakly depends on pressure (Fig.
\ref{fig3}D) or indeed on fractal dimension in the range $D=2.1-2.3$. This is
the most interesting range\cite{Persson2014}. In particular, in Fig.
\ref{fig3}C, we also show PR prediction which corresponds to a non-converging
$\left[  a_{V}\right]  _{PR}\sim\zeta^{1/3}$ for $D=2.2$ (see the Supp
Information for mathematical details) and hence results in arbitrarily large
error for large magnifications (an order of magnitude is likely). This depends
on the factor $\left\langle d^{1/3}\right\rangle /\left\langle d\right\rangle
$ which has a weaker dependence on magnification than what they assume with
their estimate on the mean diameter of the contact area $d_{rep}$.

\subsection{Criterion for stickiness}

Starting again from (\ref{area-new-formula}), and assuming the independence on
pressure of the quantity $a_{V}$, we find that both repulsive mean pressure
and adhesive mean pressure are proportional to the repulsive contact area,
which can therefore be grouped as
\begin{equation}
\frac{p_{ext}}{E^{\ast}}=\frac{p_{rep}}{E^{\ast}}-\frac{\sigma_{0}}{E^{\ast}%
}\frac{A_{ad}}{A_{nom}}=\frac{A_{rep}}{A_{nom}}\frac{\sqrt{2m_{2}}}{2}\left[
1-\frac{l_{a}}{\epsilon}\frac{3}{2}\frac{a_{V}}{q_{0}h_{rms}}\left(
\frac{\epsilon}{h_{rms}}\right)  ^{2/3}\right]  \label{eqt_pext}%
\end{equation}
where we used the identity $l_{a}/\epsilon=\sigma_{0}/E^{\ast}$ and the actual
value of the factor correlating the relative contact area $A_{rep}/A_{nom}$
with the external pressure $p_{ext}$ (see, for example, \cite{Putignano2012})
instead of the factor $2/\sqrt{\pi}$ of the original Persson's theory.

If we define the slope of the repulsive area vs. external pressure as
$A_{rep}=k\ p_{ext}/E^{\ast}$ clearly in the fractal limit because of
$m_{2}\rightarrow\infty$ in (\ref{eqt_pext}), for $\zeta\rightarrow\infty$.
Instead, we define "slope" rather as $\frac{1}{\kappa}=\frac{p_{ext}/\left(
E^{\ast}\sqrt{2m_{2}}\right)  }{A_{rep}/A_{nom}}$, and
\begin{equation}
\frac{1}{\kappa}=\frac{1}{\kappa_{rep}}-\frac{1}{\kappa_{ad}}=\frac{1}%
{2}-\frac{3a_{V}}{4q_{0}h_{rms}}\frac{l_{a}}{\epsilon}\left(  \frac{\epsilon
}{h_{rms}}\right)  ^{2/3}%
\end{equation}
and then stickiness is obtained when $1/\kappa<0$ leading to the suggested
criterion
\begin{equation}
\frac{\epsilon}{l_{a}}\left(  \frac{h_{rms}}{\epsilon}\right)  ^{2/3}<\frac
{3}{2}\frac{a_{V}}{q_{0}h_{rms}} \label{VIOLANO}%
\end{equation}
In particular, neglecting the weak dependences on pressure, magnification and
fractal dimension, we can take from the results in Fig. \ref{fig3}
$a_{V}\simeq3$ and rewrite the criterion (\ref{VIOLANO}) as%
\begin{equation}
\frac{h_{rms}}{\epsilon}<\left(  \frac{9}{4}\frac{l_{a}/\epsilon}{\epsilon
q_{0}}\right)  ^{3/5} \label{VIOLANO3}%
\end{equation}
As it can be seen eq. (\ref{VIOLANO3}) does not depend \textit{at all} on
local slopes $h_{rms}^{\prime}$ and curvatures $h_{rms}^{\prime\prime}$, hence
truncation of the PSD roughness, but \textit{only} on RMS\ amplitude of
roughness $h_{rms}$ and the largest wavelength in the roughness $q_{0}$ (both
concepts which are completely absent in PR criterion).\ 

\begin{figure}[ptbh]
\begin{center}
\includegraphics[width=17.0cm]{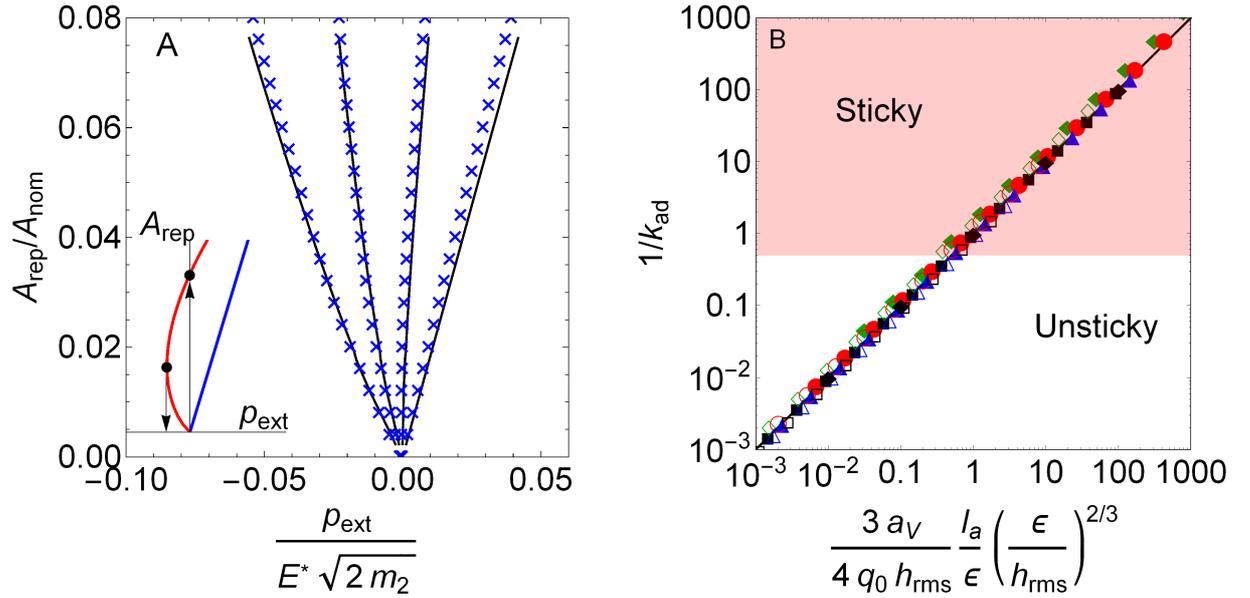}
\end{center}
\caption{(A) Contact area vs external pressure as obtained by BEM numerical
calculations (solid lines) and Persson theory (cross). The inset shows a
typical loading cycle for sticky surfaces: under zero external pressure, a
finite repulsive contact area is reached. Unloading under load control leads
to pull-off for non zero contact area. (B) Comparison between numerical
estimated $1/\kappa_{ad}$ as obtained from BEM numerical simulations and our
analytical estimate $\frac{1}{\kappa_{ad}}=$ $\frac{3a_{V}}{4q_{0}h_{rms}%
}\frac{l_{a}}{\epsilon}\left(  \frac{\epsilon}{h_{rms}}\right)  ^{2/3}$. Open
symbols for $\zeta=64$, $h_{rms}=5$ $\mu$m, while closed symbols for
$\zeta=128$, $h_{rms}=0.1$ $\mu$m, range of attraction $\varepsilon
/h_{rms}=\left[  1,1/10,1/50,1/100\right]  $ respectively diamond, circle,
triangle, square. We used $a_{V}=2.8,$ $H=0.8$ and $q_{0}=2\pi/L$ with $L=100$
$\mu$m. The shaded area identifies the region where randomly rough surfaces
are expected to be sticky, i.e. $1/\kappa_{ad}>1/2.$}%
\label{fig4}%
\end{figure}

Fig. \ref{fig4}A shows some comparison with BEM numerical calculations (solid
lines) and Persson theory (cross). In particular, the inset shows a typical
loading cycle for sticky surfaces, which shows why PR could obtain only
vertical lines in their numerical simulations in the sticky range: under zero
external pressure, upon loading the surfaces jump into contact and reach a
finite repulsive contact area. In our case, we obtain the entire curve and
hence Fig. \ref{fig4}B shows the estimated adhesive slope $1/\kappa_{ad}$ fits
well the data obtained from BEM numerical simulations over six order of
magnitude and in a wide range of parameters. The shaded area identifies the
region where randomly rough surfaces are expected to be sticky, i.e.
$1/\kappa_{ad}>1/2.$

The large wavelength wavevector cutoff of roughness $q_{0}$, which for our
scopes was defined for a pure power law fractal PSD spectrum, could in
principle take arbitrarily low values for large surfaces, which would result
in increasingly loose boundary of stickiness. For example, for a flat surface
not of micrometer size as in PR\ simulation, but of $mm$ size, we have
$\epsilon q_{0}\sim10^{-6}$ and our criterion gives $h_{rms}/\epsilon
\lesssim1000$, so the actual stickiness could persist even for roughness three
orders of magnitude larger than the range of attraction, i.e. on the order of
nearly one micron. With roughness of meters size $h_{rms}/\epsilon
\lesssim67000$ and for $km$ size, $h_{rms}/\epsilon\lesssim10^{6}$ which means
almost $mm$ amplitude of roughness. Of course these extrapolations will have
some limitation on the concept of the ideally flat surface with a pure power
law PSD of roughness. But clearly, this concepts about stickiness are
qualitatively new, and completely different from those of FT and PR.

\begin{figure}[ptbh]
\begin{center}
\includegraphics[width=17.0cm]{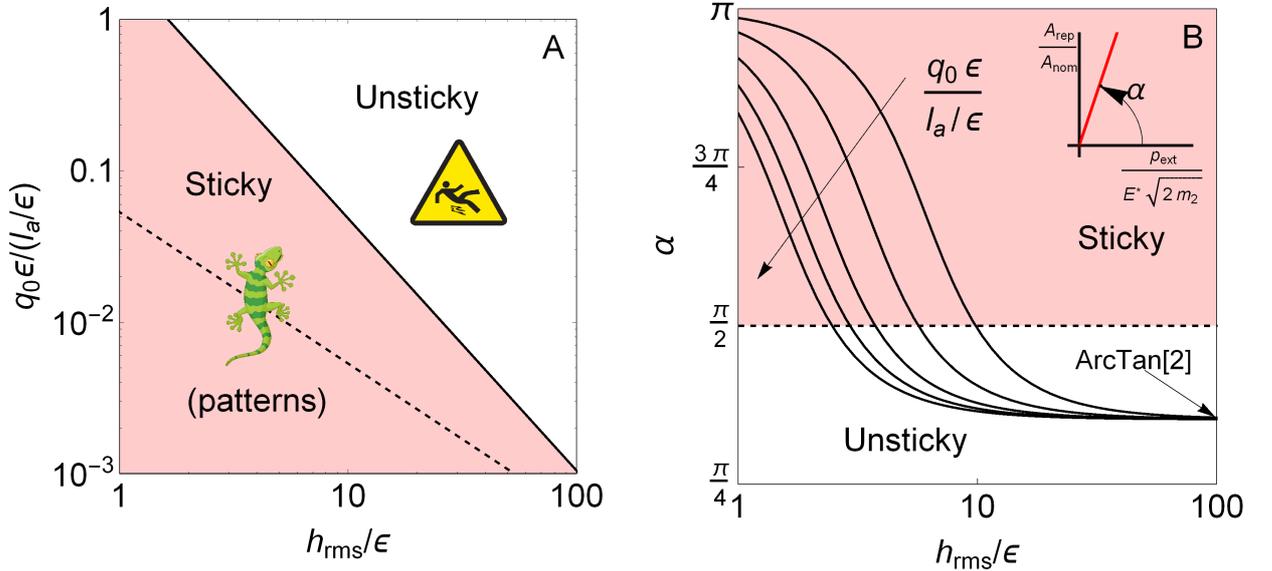}
\end{center}
\caption{(A) Adhesion map for multiscale rough surfaces according to the
present criterion (\ref{VIOLANO3}). In the plane $\frac{\epsilon q_{0}}
{l_{a}/\epsilon}$ vs $\frac{h_{rms}}{\epsilon}$\ the sticky region is shaded.
According to JTB analysis in the region below the dashed black line pattern
formation is expected with possible reduction of stickiness. (B) The slope
angle $\alpha=\arctan\left(  \kappa\right)  $ is plotted as a function of
$h_{rms}/\epsilon$ for varying $\frac{\epsilon q_{0}}{l_{a}/\epsilon}=\left[
1,3/4,1/2,1/4,1/10\right]  $ and $a_{V}=3.$}%
\label{fig5}%
\end{figure}

From JTB's results, we know that complex instabilities and patterns form at
very low RMS amplitude of roughness, and hence in the sticky range, DMT type
of analysis can be expected to hold only for approximately%
\begin{equation}
\frac{h_{rms}}{\epsilon}>\frac{4}{75}\frac{l_{a}/\epsilon}{\epsilon q_{0}}
\label{limit}%
\end{equation}
It is also likely, from JTB predictions, that stickiness is lower if this
condition is violated and a more refined analysis is needed anyway, which is
outside the possibilities of both JTB and our model. Fig. \ref{fig5} shows
stickiness map obtained with the present criterion. The shaded region in Fig.
\ref{fig5}A individuates the couple $(\frac{\epsilon q_{0}}{l_{a}/\epsilon}%
$,$\frac{h_{rms}}{\epsilon}$) which would give a stickiness. Underneath the
dashed line pattern formation is expected, and our DMT model may not be
accurate. In Fig. \ref{fig5}B the slope angle $\alpha=\arctan\left(
\kappa\right)  $ is plotted as a function of $h_{rms}/\epsilon$ for varying
$\frac{\epsilon q_{0}}{l_{a}/\epsilon}$.

\section{Application to real surfaces}

Considering our result (\ref{VIOLANO3}), it is clear that to improve
stickiness, for a given range of attractive forces $\epsilon$, we need to make
$q_{0}$ as small as possible -- this is however going to increase {$h_{rms}$}
as for a power law PSD $C\left(  q\right)  =C_{0}q^{-2\left(  H+1\right)  }$,
we have {$h_{rms}\simeq\sqrt{\pi C_{0}/H}q_{0}^{-H}$. It may be useful to
rewrite the criterion in terms }of the PSD multiplier $C_{0}$ (as usual, for
$H=0.8$)%
\begin{equation}
C_{0}<\frac{0.8}{\pi}\epsilon^{4/5}q_{0}^{2/5}\left(  \frac{9}{4}\frac{l_{a}%
}{\epsilon}\right)  ^{6/5} \label{C0}%
\end{equation}

It appears clear that we need as small roughness as possible, for a given
$q_{0}$ which is presumably dictated by size of the specimen up to some
extent, or by the process from which the surface originates. Also, we need to
have $l_{a}/\epsilon$ as high as possible, and this means \ obviously high
$\Delta\gamma$ and low $E^{\ast}$ (being $l_{a}=\Delta\gamma/E^{\ast}$).

Given $\Delta\gamma$ is in practice strongly reduced by contaminants and
various other effects to values of the order $\sim50mJ/m^{2}$, the only
reliable way to have high stickiness is to have very soft materials.

\begin{figure}[ptbh]
\begin{center}
\includegraphics[width=17.0cm]{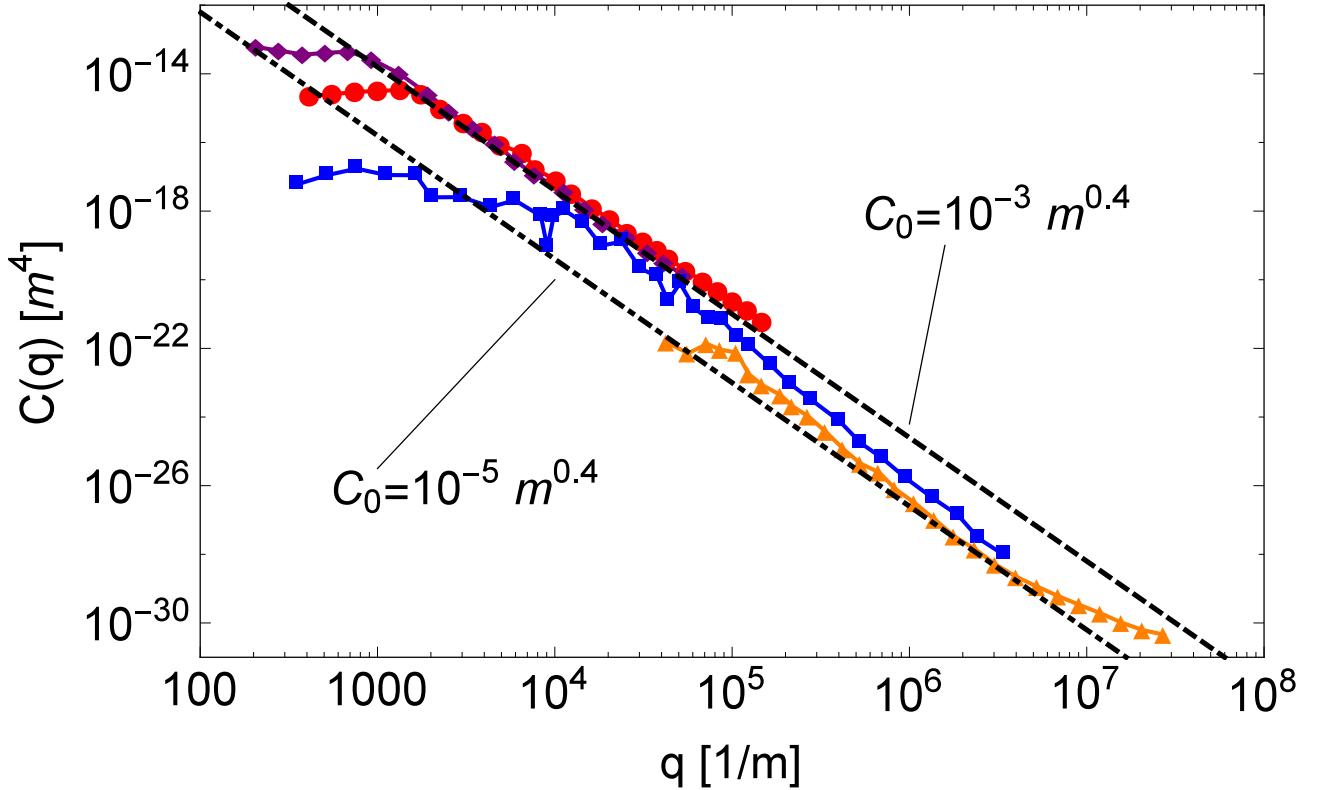}
\end{center}
\caption{PSD of typical real surfaces from (\cite{Persson2014}): blue squares
for a sandblasted PMMA, orange triangles for grinded steel, purple diamonds
and red dots are two asphalt surfaces. Black dashed line has $C_{0}=10^{-3}$
m$^{0.4},$ black dot-dashed line has $C_{0}=10^{-5}$ m$^{0.4}.$}%
\label{fig6}%
\end{figure}

As reported by Persson \cite{Persson2014}, most polished steel surfaces for
example, when measured on $L\sim0.1mm$, show $h_{rms}\sim1\mu m$. This means
$q_{0}h_{rms}\sim0.1$. This incidentally satisfies JTB condition for
"DMT"-like (\ref{limit}) practically in the entire range of possible
materials. Hence, contrary to the recent emphasys on measuring entire PSD of
surfaces, it seems therefore that for stickiness, the most important factors
are well defined, and macroscopic quantities, which are easy to measure. For a
wide range of surfaces (asphalt, sandblasted PMMA, polished steel, tape,
glass) reported in \cite{Persson2014}, see Fig. \ref{fig6}, $C_{0}$ is at most
$10^{-3}m^{0.4}$. Then, assuming $\epsilon=0.2$ $nm$ and $\Delta\gamma=50$
$mJ/m^{2}$ as a typical value, even with the uncertainty in the choice of
$q_{0}$, in the range of $q_{0}\sim10^{3}$ $\left[  m^{-1}\right]  $, our
criterion predicts $E^{\ast}<0.2$ MPa. This is for asphalt which clearly is
one of the most rough surface we can consider, and indeed outside the normal
application of PSA. Our criterion is therefore entirely compatible with the
empirical criterion by Carl Dahlquist from 3M which suggests to make tapes
only with low modulus materials $E^{\ast}<1MPa$, whose generality was so far
still scientifically unexplained. The stickiness criterion of PR is instead
further commented in the Supplementary information, and shown to have occurred
by a fitting of numerical results in a very narrow range of magnifications
$\zeta\simeq1000$ covering just from micrometer to nanometer scales (see Fig.
\ref{fig3}).

\section{Conclusions}

We have defined a new stickiness criterion, whose main factors are the low
wavevector cutoff of roughness, $q_{0}$, and the rms amplitude of roughness
$h_{rms}$. This is in striking contrast with previous theories such as Fuller
and Tabor or Pastewka and Robbins, which found a never-ending change of
stickiness with growing cutoff of PSD spectrum $\zeta$. We find that, in
principle, it is possible to have effective stickiness even with quite large
rms amplitudes, orders of magnitude larger than the range of attractive
forces. Stickiness may depend weakly on local quantities such as rms slopes
and curvatures only for narrow PSD spectra, but for realistic spectra which
are typically beyond the present brute-force simulations, the truncation of
the PSD spectrum of roughness is irrelevant.\ For robust adhesion with
different possible levels of roughness, the main characteristic affecting
stickiness is the elastic modulus, in qualitative and quantitative agreement
with Dahlquist criterion.

\section{Methods}

For BEM simulations, we use the Contact App from Pastewka (see
\cite{Pastewka2012}) with the following geometry: longest wavelength $L=100\mu
m$, $h_{rms}=0.1\mu m$, $H=0.8$, $\zeta=128$ and averaging over 8 different
realizations of the surface. There is no roll-off in the PSD and periodical
b.c. are applied for the window of size $L$. For the "precise" and broad-band
PSD spectrum, we use calculations with the Persson's contact mechanics theory,
in the version reported by Afferrante \textit{et al.}\cite{Persson2014}. Since
Persson's theory does not permit an immediate understanding of the main
parameters involved, we here derive an asymptotic theory for the attractive
area $A_{ad}$. Consider the relationship between (repulsive) contact area
ratio $\frac{A_{rep}}{A_{nom}}$ and mean pressure $p_{rep}$ which holds for
$\frac{A_{rep}}{A_{nom}}$ up to almost 30\%, where $A_{nom}$ is the nominal
contact area (original Persson's theory, \cite{Persson2001})
\begin{equation}
\frac{A_{rep}}{A_{nom}}\simeq\frac{2}{\sqrt{\pi}}\frac{p_{rep}}{\sqrt{2V}%
}=\frac{2}{\sqrt{\pi}}\frac{p_{rep}}{E^{\ast}\sqrt{m_{2}}}
\label{persson-reduced}%
\end{equation}
where $V=\frac{1}{2}E^{\ast2}m_{2}$ is the variance of full contact pressures,
and $m_{2}$ is the mean square profile slope along any direction (for a
isotropic surface). The distribution of pressures $P\left(  p\right)  $ near
the boundaries of contact (on the contact side) is at low $p$
\cite{Manners2006}%
\begin{equation}
P\left(  p\right)  \simeq\frac{pp_{rep}}{V}\sqrt{\frac{2}{\pi V}}
\label{perssonp}%
\end{equation}

Suppose the perimeter of the actual contact area [not necessarily
simply-connected] is $\Pi$. We define position on $\Pi$ by a curvilinear
coordinate $s$. In view of the asymptotic behaviour at the edge of the contact
area \cite{Johnson1985}, as noticed also by PR, we must have at every point on
$\Pi$, pressure $p$ and gap $u$ as%
\begin{equation}
p\left(  x\right)  =B\left(  s\right)  x^{1/2};\qquad u\left(  x\right)
=C\left(  s\right)  x^{3/2};\qquad
\end{equation}
where $x$ is a coordinate perpendicular to the boundary and $C\left(
s\right)  =\beta d\left(  s\right)  ^{-1/2}$ where $\beta$ is a prefactor of
order 1, $B\left(  s\right)  =\frac{3E^{\ast}\beta}{4}d\left(  s\right)
^{-1/2}$ and $d\left(  s\right)  $ is a local characteristic length scale.

It is clear that, as load is increased, existing contacts grow larger and some
new form, and the shape is very irregular. Already Greenwood and Williamson
\cite{GW} in their simple asperity theory suggested that the \textit{average}
radius of contact should remain constant with load, as a result of competition
between growing contacts and new contacts forming, and this is correctly
captured despite the strong approximations in the asperity model. In
\cite{Ciava2017} it was shown, again with a simple asperity model, but with an
exponential distribution of asperity heights, $d_{rep}$ seems indeed
completely independent on load, and that $d_{rep}=bh_{rms}^{\prime}%
/h_{rms}^{\prime\prime}$, where however $b\simeq1.63\alpha^{1/4}$ where
$\alpha$ is Nayak parameter bandwidth parameter $\alpha=m_{0}m_{4}/m_{2}%
^{2}\sim\zeta^{2H}$ and hence grows without limit with $\zeta$. So while PR
find a range of $d_{rep}$ within a factor two from $d_{rep}=4h_{rms}^{\prime
}/h_{rms}^{\prime\prime},$ we find much larger discrepancies could be found in
general for broadband PSD, i.e. larger $\zeta$ than their explored.

Hence, we shall leave the quantity $d\left(  s\right)  $ to vary arbitrarily
along the perimeter, and the probability PDF for the pressure is easily found
as%
\begin{equation}
P\left(  p\right)  =\frac{2p}{A_{nom}}\int_{\Pi}\frac{ds}{B\left(  s\right)
^{2}}=\frac{2p\Pi}{A_{nom}}I_{p} \label{p1}%
\end{equation}
where $I_{p}=\int_{0}^{1}d\widehat{s}/B\left(  \widehat{s}\right)
^{2}=\left[  4/\left(  3\beta E^{\ast}\right)  \right]  ^{2}\left\langle
d\right\rangle $, $\widehat{s}=s/\Pi$ is a normalized coordinate along the
perimeter, and $\left\langle d\right\rangle $ means the mean value of $d$. A
similar argument with the gap expression yields%
\begin{equation}
P\left(  u\right)  =\frac{1}{A_{nom}}\left(  \frac{4}{9u}\right)  ^{1/3}\Pi
I_{u} \label{p2}%
\end{equation}
where $I_{u}=\beta^{-2/3}\int_{0}^{1}d\left(  \widehat{s}\right)
^{1/3}d\widehat{s}=\beta^{-2/3}\left\langle d^{1/3}\right\rangle $.
Eliminating the perimeter from the (\ref{p1},\ref{p2}) and using
(\ref{perssonp}) results in $\Pi=\frac{pA_{nom}}{2I_{p}}\sqrt{\frac{2}{\pi
V^{3/2}}}$ and hence
\begin{equation}
P\left(  u\right)  =\ \left(  \frac{4}{9u}\right)  ^{1/3}\frac{I_{u}}{2I_{p}%
}\sqrt{\frac{2}{\pi V^{3}}}p_{rep}%
\end{equation}

Finally upon integration, we obtain (\ref{area-new-formula}), (\ref{a(V)}).

\textbf{Author contributions}. MC designed research, LA wrote the Persson code
and GV run the Persson code, while AP run the BEM code, and prepared the
figures. MC wrote the draft MS and all authors revised the MS up to its final form.

\begin{acknowledgement}
A.P. is thankful to the DFG (German Research Foundation) for funding the
projects HO 3852/11-1 and PA 3303/1-1. MC is thankful to Proff. JR\ Barber,
Costantino Creton and Steven Abbott for some discussions.
\end{acknowledgement}

\section*{Supplementary Information}

\subsection*{S1 - Comparison with PR theory}

The PR\ criterium is summarized in the following equation (eqt. 10 in their
paper)%
\begin{equation}
\frac{h_{\mathrm{rms}}^{\mathrm{\prime}}\Delta r}{2l_{a}}\left(
\frac{h_{\mathrm{rms}}^{\mathrm{\prime}}d_{rep}}{4\Delta r}\right)  ^{2/3}%
<\pi\left(  \frac{3}{16}\right)  ^{2/3}\simeq1 \label{PRcriterion}%
\end{equation}
where $d_{rep}$ is a characteristic diameter of repulsive contact areas, which
they estimate as $d_{rep}=4h_{rms}^{\prime}/h_{rms}^{\prime\prime}$,$\ $and
$h_{rms}^{\prime}$ and $h_{rms}^{\prime\prime}$\ are the rms slopes and
curvature. Finally, $\Delta r$ is the attractive range which is of the order
of atomic spacing (very close to $\epsilon$ for the Lennard-Jones potential).
When the condition (\ref{PRcriterion}) is satisfied, the surfaces in mutual
contact are suggested to be "sticky" and a finite value of the pull-off force
should occur.\bigskip\ As we have seen in the introduction, in the fractal
limit $\zeta\rightarrow\infty$, PR predict (paradoxically) stickiness for most
engineering made and natural surfaces which for Persson \cite{Persson2014}
have $H=0.8\pm0.15$. PR did not incur in this paradox since they argued in
their eqt. 9 that atomic steps limit imply $h_{rms}^{\prime\prime}%
\simeq\left(  h_{rms}^{\prime}\right)  ^{2}/\epsilon$ which simplifies their
criterion as $l_{a}/\epsilon>h_{\mathrm{rms}}^{\mathrm{\prime}}/2$. However,
in turn this condition becomes extremely difficult to satisfy or even to
check, since it is hard to define the $h_{\mathrm{rms}}^{\mathrm{\prime}}$ at
atomic scale, where surfaces having finite steps of atoms, see \cite{Luan}.

Comparison with the PR results is best done considering the attractive area
(their eqt. 6), reading%
\begin{align*}
A_{ad}  &  =A_{rep}\left(  \frac{16}{9\pi}\right)  ^{-1/3}\left(  \frac
{\pi\Delta r}{h_{rms}^{\prime}d_{rep}}\right)  ^{2/3}\\
&  =2\left(  \frac{16}{9\pi}\right)  ^{-1/3}p\frac{A_{nom}}{h_{rms}^{\prime
}E^{\ast}}\left(  \frac{\pi a_{0}\sqrt{24l_{a}/\epsilon}h_{rms}^{\prime\prime
}}{4h_{rms}^{\prime2}}\right)  ^{2/3}%
\end{align*}
where we used their equation $A_{rep}=2pA_{nom}/\left(  h_{rms}^{\prime
}E^{\ast}\right)  $ and $\Delta r/\epsilon=\sqrt{24l_{a}/\epsilon}$ from
Supplementary Information of PR paper so that for $l_{a}/\epsilon=0.05$.
Hence, it can be shown that, compared to our \ (\ref{area-new-formula}), their
estimate corresponds to
\begin{equation}
\left[  a_{V}\left(  \zeta\right)  \right]  _{PR}=1.4622q_{0}h_{rms}\left(
\frac{h_{rms}^{\prime\prime2}}{h_{rms}^{\prime7}}\right)  ^{1/3}h_{rms}^{2/3}%
\end{equation}

Now for power law tail of the PSD $C\left(  q\right)  =C_{0}q^{-2\left(
H+1\right)  }$, estimating $h_{rms}^{\prime}=\sqrt{2m_{2}},h_{rms}%
^{\prime\prime}=\sqrt{8m_{4}/3}$ (see Suppl.Info S2) we obtain as for $H=0.8$
\begin{equation}
\left[  a_{V}\left(  \zeta\right)  \right]  _{PR}=0.252\zeta^{1/3}%
\end{equation}
which results in a magnification dependence much stronger than the real
dependence we find (see Fig. \ref{fig3}C). Hence, the comparison show that
their prediction happen fortuitously approximately true only for a very narrow
range of $\zeta$. Indeed, PR's numerical results are themselves affected by
numerical errors both in the statistics of the surface, and on the level of
discretization: with a mesh of atoms of spacing $a_{0}$ and contact diameters
which are a fraction of smallest wavelength $\lambda_{s}=$\thinspace
$4...64a_{0}$, we have only few atoms to describe the contact area. However,
at very high but realistic $\zeta$, the difference grows arbitrarily large
since the correct value converges, while the PR estimate continues to grow
suggesting stickiness in all cases. The largely smaller area of adhesion was
confirmed already by independent assessment \cite{Violano2019} with yet other models.

\subsection*{S2 - On random process theory}

Assume the surface $h\left(  x,y\right)  $ has a continuous noise spectrum in
two dimensions and is described by a Gaussian stationary process. In such
case, we write%
\begin{equation}
h\left(  x,y\right)  =\sum_{n}C_{n}\cos\left[  q_{x,n}x+q_{y,n}y+\phi
_{n}\right]  \label{h(x,y)}%
\end{equation}
where the wave-components $q_{x,n}$ and $q_{y,n}$ are supposed densely
distributed throughout the $\left(  qx,qy\right)  $ plane. The random phases
$\phi_{n}$ are uniformoly distributed in the interval $[0,$ $2\pi)$. The
amplitudes $C_{n}$ are also random variables such that in any element
$dq_{x}dq_{y}$
\begin{equation}
\sum_{n}\frac{1}{2}C_{n}^{2}=C\left(  q_{x},q_{y}\right)  dq_{x}dq_{y}.
\label{Cn}%
\end{equation}

The function $C\left(  q_{x},q_{y}\right)  $ is the Power Spectral Density
(PSD) of the surface $h$, whose mean-square value can been calculated as
\begin{equation}
m_{00}=\int\int_{-\infty}^{+\infty}C\left(  q_{x},q_{y}\right)  dq_{x}dq_{y}%
\end{equation}

For isotropic roughness, using Nayak \cite{Nayak1971} definitions for the
surface
\begin{equation}
m_{rs}=\int\int_{-\infty}^{\infty}C\left[  q_{x},q_{y}\right]  q_{x}^{r}%
q_{y}^{s}dq_{x}dq_{y}%
\end{equation}
where $m_{00}$ is by definition $h_{rms}^{2}$. It can be shown by defining the
PSD and the ACF (autocorrelation function) of the partial derivatives of $h$
with respect to $x$ and $y$ coordinates, and using a relationship with the PSD
of the surface, that the above spectral moments are (see \cite{LH1957})
\begin{equation}%
\begin{tabular}
[c]{l}%
$\left\langle \left(  \frac{\partial^{r+s}h}{\partial x^{r}\partial y^{s}%
}\right)  ^{2}\right\rangle =m_{2r,2s}$\\
$\left\langle \left(  \frac{\partial^{r+s}h}{\partial x^{r}\partial y^{s}%
}\right)  ^{2}\right\rangle =\left(  -1\right)  ^{\frac{1}{2}\left(
r+s-r^{\prime}-s^{\prime}\right)  }m_{r+r^{\prime},s+s^{\prime}}\quad$%
or$\quad0,$%
\end{tabular}
\ \
\end{equation}
depending on $\left(  s+r-r^{\prime}-s^{\prime}\right)  $ is even or odd.

Nayak finds for isotropic surface,
\begin{align}
m_{20}  &  =m_{02}=m_{2};\qquad m_{11}=m_{13}=m_{31}=0\nonumber\\
m_{00}  &  =m_{0};\qquad3m_{22}=m_{40}=m_{04}=m_{4}%
\end{align}
meaning when there is no second subscript the \textit{profile statistics for
isotropic surface}, which is independent on the direction chosen.

For slopes, with the common definition of their rms value is (also used by PR)%
\begin{equation}
h_{rms}^{\prime}=\sqrt{\left\langle \left\vert \nabla h\right\vert
^{2}\right\rangle }=\sqrt{\left\langle \left(  \frac{\partial h}{\partial
x}\right)  ^{2}+\left(  \frac{\partial h}{\partial y}\right)  ^{2}%
\right\rangle }=\sqrt{2m_{2}}%
\end{equation}
where the equality depends on the result that, for an isotropic surface, the
orthogonal components $\frac{\partial h}{\partial x}$ and $\frac{\partial
h}{\partial y}$ are uncorrelated.

The definition of RMS curvature $h_{rms}^{\prime\prime}$ is less common, but
we shall follow PR in defining
\begin{align}
h_{rms}^{\prime\prime}  &  =\sqrt{\left\langle \left(  \nabla^{2}h\right)
^{2}\right\rangle }=\sqrt{\left\langle \left(  \frac{\partial^{2}h}{\partial
x^{2}}\right)  ^{2}+\left(  \frac{\partial^{2}h}{\partial y^{2}}\right)
^{2}+2\left(  \frac{\partial^{2}h}{\partial x^{2}}\right)  \left(
\frac{\partial^{2}h}{\partial y^{2}}\right)  \right\rangle }\nonumber\\
&  =\sqrt{m_{40}+m_{04}+2m_{22}}=\sqrt{8m_{4}/3}%
\end{align}

PR do not recur to asperity theories to define the mean radius involved in the
contact areas $R$ , which is defined for asperities by Nayak's theory
\cite{Nayak1971}. In this case, $R$ would vary between $1/\left(  2\sqrt
{m_{4}}\right)  $ at low bandwidths and $1/\left(  4.73\sqrt{m_{4}}\right)
\ $at high bandwidths, a change of a factor $2.3$. PR rather estimate $R$ from
the rms curvature which indeed results in $R=2/h_{rms}^{\prime\prime}%
=1/\sqrt{2m_{4}/3}$ as it can be easily verified, and which, incidentally, is
larger than the entire range expected from Nayak's analysis. Using the Nayak
estimate however would only improve the results in some range as it would
increase the area of attraction, but in the fractal limit, it would also not
work correctly. In their limited band of investigation, PR found $d_{rep}$ to
be always within a factor of $2$ of their estimate, but this factor may
largely change were they to consider broader band of roughness.
\end{document}